\documentclass[12 pt]{article}
\pdfoutput=1 
\usepackage{amsmath,mathrsfs}
\usepackage{amssymb}
\allowdisplaybreaks
\usepackage{amscd}
\usepackage{graphicx}
\usepackage{tabularx,booktabs}
\usepackage{array}
\newcolumntype{Z}{>{\centering\let\newline\\\arraybackslash\hspace{0pt}}X}
\usepackage[footnotesize,bf,textfont={it},margin=1 cm]{caption}

\def\be{\begin{equation}}
\def\ee{\end{equation}}
\def\frc#1#2{\relax\ifmmode{\textstyle\frac{#1}{#2}} 
                    \else$\frac{#1}{#2}$\fi}         

\def\rI{{{}_{\rm I}}}
\def\rJ{{{}_{\rm J}}}

\def\hk{{\hat k}}
\def\hi{{\hat\imath}}

\def\fracm#1#2{\hbox{\large{${\frac{{#1}}{{#2}}}$}}}

\def\vCent#1{\vcenter{\hbox{\hss#1\hss}}}

\def\pp{{\mathchoice
          {
              \kern 1pt%
              \raise 1pt
              \vbox{\hrule width5pt height0.4pt depth0pt
                    \kern -2pt
                    \hbox{\kern 2.3pt
                          \vrule width0.4pt height6pt depth0pt
                          }
                    \kern -2pt
                    \hrule width5pt height0.4pt depth0pt}%
                    \kern 1pt
           }
            {
              \kern 1pt%
              \raise 1pt
              \vbox{\hrule width4.3pt height0.4pt depth0pt
                    \kern -1.8pt
                    \hbox{\kern 1.95pt
                          \vrule width0.4pt height5.4pt depth0pt
                          }
                    \kern -1.8pt
                    \hrule width4.3pt height0.4pt depth0pt}%
                    \kern 1pt
            }
            {
              \kern 0.5pt%
              \raise 1pt
              \vbox{\hrule width4.0pt height0.3pt depth0pt
                    \kern -1.9pt  
                    \hbox{\kern 1.85pt
                          \vrule width0.3pt height5.7pt depth0pt
                          }
                    \kern -1.9pt
                    \hrule width4.0pt height0.3pt depth0pt}%
                    \kern 0.5pt
            }
            {
              \kern 0.5pt%
              \raise 1pt
              \vbox{\hrule width3.6pt height0.3pt depth0pt
                    \kern -1.5pt
                    \hbox{\kern 1.65pt
                          \vrule width0.3pt height4.5pt depth0pt
                          }
                    \kern -1.5pt
                    \hrule width3.6pt height0.3pt depth0pt}%
                    \kern 0.5pt
            }
        }}

\def\mm{{\mathchoice
                       {
                             \kern 1pt
               \raise 1pt    \vbox{\hrule width5pt height0.4pt depth0pt
                                  \kern 2pt
                                  \hrule width5pt height0.4pt depth0pt}
                             \kern 1pt}
                       {
                            \kern 1pt
               \raise 1pt \vbox{\hrule width4.3pt height0.4pt depth0pt
                                  \kern 1.8pt
                                  \hrule width4.3pt height0.4pt depth0pt}
                             \kern 1pt}
                       {
                            \kern 0.5pt
               \raise 1pt
                            \vbox{\hrule width4.0pt height0.3pt depth0pt
                                  \kern 1.9pt
                                  \hrule width4.0pt height0.3pt depth0pt}
                            \kern 1pt}
                       {
                           \kern 0.5pt
             \raise 1pt  \vbox{\hrule width3.6pt height0.3pt depth0pt
                                  \kern 1.5pt
                                  \hrule width3.6pt height0.3pt depth0pt}
                           \kern 0.5pt}
                       }}

\def\ad{{\kern0.5pt
                   \alpha \kern-5.05pt \raise5.8pt\hbox{$\textstyle.$}\kern
0.5pt}}

\def\bd{{\kern0.5pt
                   \beta \kern-5.05pt \raise5.8pt\hbox{$\textstyle.$}\kern
0.5pt}}

\def\qd{{\kern0.5pt
                   q \kern-5.05pt \raise5.8pt\hbox{$\textstyle.$}\kern
0.5pt}}
\def\Dot#1{{\kern0.5pt
     {#1} \kern-5.05pt \raise5.8pt\hbox{$\textstyle.$}\kern
0.5pt}}

\catcode`@=11
\def\un#1{\relax\ifmmode\@@underline#1\else
        $\@@underline{\hbox{#1}}$\relax\fi}
\catcode`@=12

\def\a{\alpha}
\def\b{\beta}

\def\e{\epsilon}

\def\l{\lambda}
\def\m{\mu}

\def\r{\rho}
\def\s{\sigma}
\def\t{\tau}

\def\dslash{\not{\hbox{\kern-2pt $\partial$}}}
\def\Dslash{\not{\hbox{\kern-4pt $D$}}}
\def\pslash{\not{\hbox{\kern-2.3pt $p$}}}
 \newtoks\slashfraction
 \slashfraction={.13}
 \def\slash#1{\setbox0\hbox{$ #1 $}
 \setbox0\hbox to \the\slashfraction\wd0{\hss \box0}/\box0 }
 
\font\ro=cmsy10                          
\def\kcr{{\hbox{\ro \char'170}}}                
\def\ktl{{\hbox{\ro \char'170}}}        
\def\ktr{{\hbox{\ro \char'170}}}        
\def\kbl{{\hbox{\ro \char'170}}}        
\def\kbr{{\hbox{\ro \char'170}}}        

\def\plpl{\raise-2pt\hbox{$\raise3pt\hbox{$_+$}\hskip-6.67pt\raise0.0pt
\hbox{$^+$}\hskip 0.01pt$}}
\def\mimi{\raise-2pt\hbox{$\raise3pt\hbox{$_-$}\hskip-6.67pt\raise0.0pt
\hbox{$^-$}\hskip 0.01pt$}} 

\def\bo{{\raise.15ex\hbox{\large$\Box$}}}               
\def\pa{\partial}                                       
\def\TH{{\raise.2ex\hbox{$\displaystyle \bigodot$}\mskip-4.7mu \llap H \;}}
\def\face{{\raise.2ex\hbox{$\displaystyle \bigodot$}\mskip-2.2mu \llap {$\ddot
        \smile$}}}                                      

\def\Tilde#1{\widetilde{#1}}                    
\def\Hat#1{\widehat{#1}}                        
\def\Bar#1{\overline{#1}}                       
\def\leftrightarrowfill{$\mathsurround=0pt \mathord\leftarrow \mkern-6mu
        \cleaders\hbox{$\mkern-2mu \mathord- \mkern-2mu$}\hfill
        \mkern-6mu \mathord\rightarrow$}
\def\dvec#1{\vbox{\ialign{##\crcr
        \leftrightarrowfill\crcr\noalign{\kern-1pt\nointerlineskip}
        $\hfil\displaystyle{#1}\hfil$\crcr}}}           
\def\dt#1{{\buildrel {\hbox{\LARGE .}} \over {#1}}}     

\def\fracm#1#2{\hbox{\large{${\frac{{#1}}{{#2}}}$}}}
\def\frac#1#2{{\textstyle{#1\over\vphantom2\smash{\raise.20ex
        \hbox{$\scriptstyle{#2}$}}}}}                   
\def\sfrac#1#2{{\vphantom1\smash{\lower.5ex\hbox{\small$#1$}}\over
        \vphantom1\smash{\raise.4ex\hbox{\small$#2$}}}} 
\def\bfrac#1#2{{\vphantom1\smash{\lower.5ex\hbox{$#1$}}\over
        \vphantom1\smash{\raise.3ex\hbox{$#2$}}}}       
\def\afrac#1#2{{\vphantom1\smash{\lower.5ex\hbox{$#1$}}\over#2}}    
\def\on#1#2{\mathop{\null#2}\limits^{#1}}               

\def\pa{\partial}      
\newcommand{\bm}[1]{\mbox{\boldmath$#1$}}

\def\ad{{\dot{\alpha}}}
\def\bd{{\dot{\beta}}}

\font\ro=cmsy10                          
\def\kcr{{\hbox{\ro \char'170}}}                
\def\ktl{{\hbox{\ro \char'170}}}        
\def\ktr{{\hbox{\ro \char'170}}}        
\def\kbl{{\hbox{\ro \char'170}}}        
\def\kbr{{\hbox{\ro \char'170}}}        

\parskip=\medskipamount                 
\thicklines                         

\def\border{                                            
        \setlength{\unitlength}{1mm}
        \newcount\xco
        \newcount\yco
        \xco=-21
        \yco=12
        \begin{picture}(140,0)
        \put(\xco,\yco){$\ktl$}
        \advance\yco by-1
        {\loop
        \put(\xco,\yco){$\kcr$}
        \advance\yco by-2
        \ifnum\yco>-240
        \repeat
        \put(\xco,\yco){$\kbl$}}
        \xco=158
        \yco=12
        \put(\xco,\yco){$\ktr$}
        \advance\yco by-1
        {\loop
        \put(\xco,\yco){$\kcr$}
        \advance\yco by-2
        \ifnum\yco>-240
        \repeat
        \put(\xco,\yco){$\kbr$}}
        \put(-20,13){\tiny **University of Maryland * Center for String and
         Particle  Theory* Physics Department***University of Maryland *Center
        for String and Particle  Theory** }
        \put(-20,-241.5){\tiny **University of Maryland * Center for String and
         Particle  Theory* Physics Department***University of Maryland *Center
        for String and Particle  Theory** }
        \end{picture}
        \par\vskip-8mm}

\def\headpic{                                           
        \indent
        \setlength{\unitlength}{.4mm}
        \thinlines
        \par
        \begin{picture}(29,16)
        \put(165,16){\line(1,0){4}}
        \put(170,16){\line(1,0){4}}
        \put(180,16){\line(1,0){4}}
        \put(175,0){\line(1,0){4}}
        \put(180,0){\line(1,0){4}}
        \put(185,0){\line(1,0){4}}
        \put(169,0){\line(0,1){16}}
        \put(170,0){\line(0,1){16}}
        \put(179,0){\line(0,1){16}}
        \put(180,0){\line(0,1){16}}
        \put(184,0){\line(0,1){16}}
        \put(185,0){\line(0,1){16}}
        \put(169,16){\oval(8,32)[bl]}
        \put(170,16){\oval(8,32)[br]}
        \put(179,0){\oval(8,32)[tl]}
        \put(185,0){\oval(8,32)[tr]}
        \end{picture}
        \par\vskip-6.5mm
        \thicklines}
\def\endtitle{\end{quotation}\newpage}                  

\newskip\humongous \humongous=0pt plus 1000pt minus 1000pt
\def\caja{\mathsurround=0pt}
\def\eqalign#1{\,\vcenter{\openup2\jot \caja
        \ialign{\strut \hfil$\displaystyle{##}$&$
        \displaystyle{{}##}$\hfil\crcr#1\crcr}}\,}
\newif\ifdtup

\begin{document}

\def\dt#1{\on{\hbox{\bf .}}{#1}}                
\def\Dot#1{\dt{#1}}

\def\gg{{\hbox{\sc g}}}
\border\headpic {\hbox to\hsize{
\hfill
{UMDEPP-015-011}}}
\par \noindent
{ \hfill
}
\par

\setlength{\oddsidemargin}{0.3in}
\setlength{\evensidemargin}{-0.3in}
\begin{center}
{\large\bf A Lorentz Covariant Holoraumy-Induced 
``Gadget''  \\[.1in] 
From Minimal Off-Shell  
4D, $\cal N$ = 1 Supermultiplets}\\[.4in]
S.\, James Gates, Jr.\footnote{gatess@wam.umd.edu}${}^{\dagger}$,
Tyler Grover\footnote{tgrover@terpmail.umd.edu}${}^{\dagger}$, 
Miles David Miller-Dickson\footnote{milesmdickson@yahoo.com}${}^{\dagger}$, 
Benedict A. Mondal\footnote{benedict.mondal@gmail.com}${}^{\dagger}$,
Amir Oskoui\footnote{amir.osko@gmail.com}${}^{\dagger}$,
Shirash Regmi\footnote{c$_-$rash2005@hotmail.com}${}^{\dagger}$,
Ethan Ross\footnote{eross@ualberta.ca}${}^{\dagger}{}^{*}$, and
Rajath Shetty\footnote{rajathks@outlook.com}${}^{\dagger}$
\\[0.2in]
${}^\dag${\it Center for String and Particle Theory\\
Department of Physics, University of Maryland\\
College Park, MD 20742-4111 USA}
\\[0.1in] 
and \\[0.1in] 
 ${}^*${\it Department of Mathematics\\
University of Alberta\\
Edmonton, AB 3 Canada T6G 2R3}
\\[0.2in] 
{\bf ABSTRACT}\\[.01in]
\end{center}
\begin{quotation}
{Starting from three minimal off-shell 4D, $\cal N$ = 1 supermultiplets,  
using constructions solely defined within the confines of the four 
dimensional field theory we show the existence of a ``gadget'' - a 
member of a class of metrics on the representation space of the 
supermultiplets - whose values directly and completely correspond 
to the values of a metric defined on the 1d, $N$ = 4 adinkra networks 
adjacency matrices corresponding  to the projections of the four 
dimensional supermultiplets.}
\\[0.3in]
\vfill
\endtitle

\setcounter{equation}{0}
\section{Introduction}
$~~~$
The main purpose of this work is to extend a mathematical gadget previously
found to exist in the realm of supersymmetrical quantum mechanics models
\cite{KIAS,KIAS2,adnkholor,KIAS3} into four dimensional 
field theory with simple supersymmetry.  This will provide a new example
of ``SUSY Holography'' \cite{ENUF7} - SUSY QM can realize aspects of SUSY
QFT.

As we shall show in chapter two, it is possible (within purely four dimensional 
supersymmetrical field theories) to uncover the existence of a ``Lorentz covariant 
fermionic holoraumy tensor'' similar to that discovered within one dimensional 
models \cite{KIAS,KIAS2,adnkholor,KIAS3} {\em {but with no reference 
whatsoever to lower dimensional constructs}}.  In turn this permits the definition 
of a mathematical gadget that relates to the properties of adjacency matrices 
\cite{ENUF3,ENUF4} of bipartite graphs (given the name `{\em {adinkras}}')
\cite{adnk1} - \cite{Top&3}.  This is similar to the results of \cite{adnkColor}, 
wherein it was shown that a certain parameter ($\chi{}_{\rm o}$ - initially 
discovered in the study of four-color adinkra networks) can be defined solely 
using concepts from 4D, $\cal N$ = 1 superfield theory constructions.  

We will next review the construction of the gadget for 1d supermultiplets based 
on adinkra networks in the following chapter.  This discussion in the past has  
been shown to lead to a metric on the representation space of adinkra 
networks.  

We follow this with a discussion containing a plausibility argument for why the 
commutator (as opposed to the anti-commutator) of supercharges provides 
an appropriate starting point in discussions of a representation space metric.  

We conclude with a summary and observations and include one appendix with details
useful for some of the calculations.

\section{A 4D, $\cal N $ = 1 Minimal Supermultiplet Gadget} 
$~~~$
In this chapter, we wish to perform some calculations solely  within the context 
of off-shell 4D, $\cal N $ = 1 minimal supermultiplets.  As is well known, there 
are essentially three such supermultiplets\footnote{For the purposes of our 
discussion, we will here ignore the existence of variant  \newline $~~~\,~~$ 
representations and parity reflected representations of these three basic ones.}; the 
chiral supermultiplet (CS), vector supermultiplet (VS), and tensor supermultiplet 
(TS).  

Each such supermultiplet contains 4 bosonic degrees of freedom and
4 fermionic degrees of freedom.  In reaching this conclusion, the degrees of 
freedom of gauge fields are counted only using their gauge-independent ones.
The supermultiplets are specified by a set of component fields and the 
action of the D${}_a$ operators as realized on the component fields according 
to the following rules. \vskip.01in
$\bm {Chiral~Supermultiplet: ~(A, \, B, \,  \psi_a , \, F, \, G)}$
\be
 \eqalign{
{~~~~} {\rm D}_a A ~&=~ \psi_a  ~~~~~~~~~~~~~~,~~~~
{\rm D}_a B ~=~ i \, (\gamma^5){}_a{}^b \, \psi_b  ~~~~~~~~~, \cr
{\rm D}_a \psi_b ~&=~ i\, (\gamma^\mu){}_{a \,b}\,  \partial_\mu A 
~-~  (\gamma^5\gamma^\mu){}_{a \,b} \, \partial_\mu B ~-~ i \, C_{a\, b} 
\,F  ~+~  (\gamma^5){}_{ a \, b} G  ~~, \cr
{\rm D}_a F ~&=~  (\gamma^\mu){}_a{}^b \, \partial_\mu \, \psi_b   
~~~,~~~ 
{\rm D}_a G ~=~ i \,(\gamma^5\gamma^\mu){}_a{}^b \, \partial_\mu \,  
\psi_b  ~~~,
} \label{QT1}
\ee \indent
$\bm {Vector~Supermultiplet:~ (A{}_{\mu} , \, \l_b , \,  {\rm d})}$
\be
\eqalign{
{~~~~} {\rm D}_a \, A{}_{\mu} ~&=~  (\gamma_\mu){}_a {}^b \,  \l_b  ~~~, \cr
{\rm D}_a \l_b ~&=~   - \,i \, \fracm 14 ( [\, \gamma^{\mu}\, , \,  \gamma^{\nu} 
\,]){}_a{}_b \, (\,  \partial_\mu  \, A{}_{\nu}    ~-~  \partial_\nu \, A{}_{\mu}  \, )
~+~  (\gamma^5){}_{a \,b} \,    {\rm d} ~~,  {~~~~~~~}  \cr
{\rm D}_a \, {\rm d} ~&=~  i \, (\gamma^5\gamma^\mu){}_a {}^b \, 
\,  \partial_\mu \l_b  ~~~, \cr
}  \label{QT2}
\ee
 \indent
$\bm {Tensor~Supermultiplet: ~(\varphi, \, B{}_{\mu \, \nu }, \,  \chi_a )}$
\be
 \eqalign{
{\rm D}_a \varphi ~&=~ \chi_a  ~~~,~~~~~ 
{\rm D}_a B{}_{\mu \, \nu } ~=~ -\, \fracm 14 ( [\, \gamma_{\mu}
\, , \,  \gamma_{\nu} \,]){}_a{}^b \, \chi_b  ~~~, \cr
{\rm D}_a \chi_b ~&=~ i\, (\gamma^\mu){}_{a \,b}\,  \partial_\mu \varphi 
~-~  (\gamma^5\gamma^\mu){}_{a \,b} \, \e{}_{\mu}{}^{\r \, \s \, \t}
\partial_\r B {}_{\s \, \t}~~. {~~~~~~~~~~~~~~\,~~}
}  \label{QT3}
\ee
Up to gauge transformations, these satisfy the equation
\be
 \eqalign{  {~~~~~}
\{ ~ {\rm D}_a  \,,\,  {\rm D}_b ~\} \, 
~&=~  i\, 2 \, (\gamma^\mu){}_{a \,b}\,  \partial_\mu \,   ~~~,
}
\ee
when calculated upon each component field.

However, it is also possible to use the results in equations (\ref{QT1}), 
(\ref{QT2}), and  (\ref{QT3}) to calculate instead the commutators of
the D${}_a$  operators as evaluated on each field.  We find the 
following results which extend those found in \cite{adnkholor}.

$\bm {Chiral~Supermultiplet}$
\be   \eqalign{  {~~~~~~}
[\, {\rm D}_a  ~,~ {\rm D}_b  \,] A~&=~-i\, 2 \,C{}_{a b} F + 2({\gamma^5}){
}_{a b} G -2({\gamma^5 \gamma^{\mu}}){}_{a b} \pa_{\mu} B~~~, \cr
[\, {\rm D}_a  ~,~ {\rm D}_b  \,] B~&=~ i\, 2 \,C{}_{a b} G + 2({\gamma^5}){
}_{a b} F +  2({\gamma^5 \gamma^{\mu}}){}_{a b} \pa_{\mu} A ~~~, \cr
[\, {\rm D}_a  ~,~ {\rm D}_b  \,]\psi_c~&=~ -i ({\gamma^5 \gamma^{\mu}}){
}_{a b}(\gamma^{5}[\gamma_{\mu} \,,\, \gamma^{\sigma}])_c{}^d\pa_{\sigma}\psi_d~~~, \cr
[\, {\rm D}_a  ~,~ {\rm D}_b  \,] F~&=~-i\, 2 \,C_{ab}\eta{}^{\m \sigma}\pa_{\mu}
\pa_{\sigma} A +2(\gamma^{5})_{ab}\eta{}^{\m \sigma}\pa_{\mu}\pa_{\sigma} 
B -2({\gamma^5 \gamma^{\mu}}){}_{a b} \pa_{\mu} G~~~, \cr
[\, {\rm D}_a  ~,~ {\rm D}_b  \,] G~&=~ 
i\, 2 \,C_{ab}\eta{}^{\m \sigma}\pa_{\mu}\pa_{\sigma} B  +
2(\gamma^{5})_{ab}\eta{}^{\m \sigma} \pa_{\mu}\pa_{\sigma} A 
+2({\gamma^5 \gamma^{\mu}}){}_{a b} \pa_{\mu} F~~~, \cr
}   \label{HCS}
\ee

$\bm {Vector~Supermultiplet}$
\be   \eqalign{
[\, {\rm D}_a  ~,~ {\rm D}_b  \,] A_\mu~&=~-2\epsilon^{\sigma\nu}{}_{\mu
\alpha} (\gamma^5 \gamma^{\alpha})_{ab} \pa_{\sigma} A_\nu - 2
(\gamma^5 \gamma_{\mu})_{ab} {\rm d}~~~, \cr
[\, {\rm D}_a  ~,~ {\rm D}_b  \,]{\rm d}~&=~2(\gamma^5 \gamma^{\mu})_{
ab}\partial^\nu  \left( \pa_{\mu}A_{\nu}  ~-~ \pa_{\nu}A_{\mu}      \right)~~~, \cr
[\, {\rm D}_a  ~,~ {\rm D}_b  \,]\lambda_c~&=~ -i\, 2 \, C{}_{ab}(\gamma^{\mu}
)_c{}^d \pa_{\mu} \lambda_d -i \, 2 \,(\gamma^5 ){}_{ab}(\gamma^ 5 \gamma^{
\mu})_c{}^d \pa_{\mu} \lambda_d \cr 
&{~~~~~} -i\, 2 \, (\gamma^5 \gamma^{\mu}){}_{ab} 
(\gamma^5)_c{}^d \pa_{\mu} \lambda_d ~~~,
}   \label{HVS}
\ee

$\bm {Tensor~Supermultiplet}$
\be   \eqalign{
[\, {\rm D}_a  ~,~ {\rm D}_b  \,]\varphi~&=~2(\gamma^5 \gamma^{\mu}){}_{a 
b}\epsilon^\rho{}_\mu^{\ \alpha\beta} \partial_\rho B_{\alpha\beta}  ~~~, \cr
[\, {\rm D}_a  ~,~ {\rm D}_b  \,] B_{\mu\nu}~&=~-\epsilon_{\mu\nu}{}^{\alpha
}{}_{\beta}(\gamma^5\gamma^\beta){}_{ab}\partial_\alpha \varphi+4(\gamma^5
\gamma_{[\nu}){}_{ab}\epsilon^\rho{}_{\mu]}{}^{\alpha\beta}\partial_\rho 
B_{\alpha\beta} ~~~, \cr
[\, {\rm D}_a  ~,~ {\rm D}_b  \,]\chi_c~&=~ i\, 2 \, C{}_{ab}(\gamma^{\mu})_c{
}^d \pa_{\mu}  \chi_d -i \, 2 \,(\gamma^5){}_{ab}  (\gamma^5 \gamma^{\mu
})_c{}^d \pa_{\mu} \chi_d  \cr 
&{~~~~}
+i\, 2 \,(\gamma^5 \gamma^{\mu}){}_{a b}(\gamma^5
)_c{}^d \pa_{\mu} \chi_d  ~~~. 
}   \label{HTS}
\ee

Let us further focus only on the results for the fermions in each of the 
calculations in (\ref{HCS}), (\ref{HVS}), and (\ref{HTS}).  This brings 
us to the results below.

$\bm {Chiral~Supermultiplet~Fermion}$
\be   \eqalign{
[\, {\rm D}{}_{a} ~,~ {\rm D}{}_{b} \, ] \,\psi_{c} ~&=~ -i \, (\gamma^5 \gamma^{
\nu} )_{ab}(\gamma^5 [ \, \gamma {}_{\nu} ~,~ \gamma^{\mu} \, ])_{c}{}^{d}
\pa_{\mu} \psi_{d}   {~~~~~~~~~~~~~~~~}    \cr
~&\equiv~  \left[{ {\bm H}}{}^{\mu}{}^{(CS)}\right]{}_{a \, b \, c}{}^d \, 
\left(\pa_{\mu} \psi_{d}  \, \right)  ~~~, 
} \label{HCSf} 
\ee

$\bm {Vector~Supermultiplet~Fermion}$
\be \eqalign{
[\, {\rm D}{}_{a} ~,~ {\rm D}{}_{b} \, ] \,\lambda_{c} ~&=~ -i 2\, C_{ab}
(\gamma^{\m})_c{}^{d}\pa_\mu\lambda_{d} ~-~ i 2 \, (\gamma^{5}
)_{ab} (\gamma^{5}\gamma^{\m})_{c}{}^{d} \pa_\mu \lambda_{d}    \cr
 &~~~~~-i 2 \,(\gamma^{5}\gamma^{\m})_{ab}(\gamma^{5})_{c}{
 }^{d}\pa_\mu  \lambda_d   \cr
~&\equiv~  \left[{ {\bm H}}{}^{\mu}{}^{(VS)}\right]{}_{a \, b \, c}{}^d \, 
\left(\pa_{\mu} \l_{d}  \, \right)   ~~~,
} \label{HVSf} 
\ee

$\bm {Tensor~Supermultiplet~Fermion}$
\be \eqalign{
[\, {\rm D}{}_{a} ~,~ {\rm D}{}_{b} \, ] \,\chi_{c} ~&=~ i 2 \, C_{ab}(\gamma^{\mu}
)_c{}^{d}\pa_\mu\chi_{d}~-~ i 2\, (\gamma^5)_{ab}(\gamma^5\gamma^{\mu}
)_{c}{}^{d}\pa_\mu\chi_{d}  {~~~~}   \cr   
&~~~~~+ i2\,(\gamma^5\gamma^{\mu})_{ab}( \gamma^{5} )_{c}{}^{d}
\pa_\mu\chi_d  \cr
~&\equiv~  \left[{ {\bm H}}{}^{\mu}{}^{(TS)}\right]{}_{a \, b \, c}{}^d \, 
\left(\pa_{\mu} \chi_{d}  \, \right)  ~~~. 
}  \label{HTSf}
\ee

We derive (directly from the four dimensional formulations of each of 
(CS), (VS), and (TS) cases respectively) associated quantities $\left[{ {\bm 
H}}{}^{\mu}{}^{(CS)}\right]{}_{a \, b \, c}{}^d$, $\left[{ {\bm H}}{}^{\mu}{}^{(VS)
}\right]{}_{a \, b \, c}{}^d$, and $\left[{ {\bm H}}{}^{\mu }{}^{(TS)}\right]{}_{a \, 
b \, c}{}^d$.  These are holoraumy tensors defined purely in terms of four 
dimensional Lorentz covariant concepts.  Unlike their SUSY QM analogues 
\cite{KIAS2,adnkholor,KIAS3}, these also carry a Lorentz vector index.  Since 
each of these 4D holoraumy tensors carries additional four spinor-indices, 
by performing contractions over these spinor indices we can also form 
mathematical gadgets (i.e. - proposed metrics on the representation spaces) 
similar to those discussed in the one-dimensional cases.  

Guided by our experience from working with 0-brane reductions, we require 
a Lorentz covariant gadget (denoted by ${\widehat  {\cal G}} [ ({\widehat {\cal 
R}})  ,  ({\widehat {\cal R}}^{\prime})\, ]$) defined over the four dimensional 
supermultiplet representations $({\widehat {\cal R}})$ and $({\widehat {\cal 
R}}^{\prime})$.   We propose an ansatz of the form
\be
 \eqalign{ {~~}
{\widehat  {\cal G}} [  ({\widehat {\cal R}}) , ({\widehat {\cal R}}^{\prime}) ] 
~&=~ m_1 \,  [{ {\bm H}}{}^{\mu}{}^{({\widehat {\cal R}})}
]{}_{a \, b \, c}{}^d  \, [ { {\bm H}}{}_{\mu}{}^{({\widehat {\cal 
R}}^{\prime})} ]{}^{a \, b}{}_{d}{}^c   \cr
&~~~~+~ m_2 \,   (\gamma^{\a})_c^{~e}  \,  [{ {\bm H}}{}^{\mu}{
}^{({\widehat {\cal R}})} ]{}_{a \, b \, e}{}^f  \, (\gamma_{\a})_f^{
~d} \, [ { {\bm H}}{}_{\mu}{}^{({\widehat {\cal R}}^{\prime})} ]{}^{a \, 
b}{}_{d}{}^c                \cr
&~~~~+~ m_3 \,   ([\, \gamma^{\a}  ~,~ \gamma^{\b} \,])_c^{~e} \,  
[{ {\bm H}}{}^{\mu}{}^{({\widehat {\cal R}})} ]{}_{a \, b \, 
e}{}^f  \, ([\, \gamma_{\a}  ~,~ \gamma_{\b} \,])_f^{~d} \, [ { {\bm 
H}}{}_{\mu}{}^{({\widehat {\cal R}}^{\prime})} ]{}^{a \, b}{}_{d}{}^c  
{~~~~~~}  \cr
&~~~~+~ m_4 \,    (\gamma^5 \gamma^{\a})_c^{~e}  \,  [{ {\bm 
H}}{}^{\mu}{}^{({\widehat {\cal R}})} ]{}_{a \, b \, e}{}^f  \, (\gamma^5
\gamma_{\a})_f^{~d} \, [ { {\bm H}}{}_{\mu}{}^{({\widehat {\cal R}}^{\prime
})} ]{}^{a \, b}{}_{d}{}^c    \cr
&~~~~+~ m_5 \,   (\gamma^5)_c^{~e}  \,  [{ {\bm H}}{
}^{\mu}{}^{({\widehat {\cal R}})} ]{}_{a \, b \, e}{}^f  \, (\gamma^5
)_f^{~d} \, [ { {\bm H}}{}_{\mu}{}^{({\widehat {\cal R}}^{\prime})} ]{}^{a 
\, b}{}_{d}{}^c  ~~~,
} 
\ee
and then seek to find constants $m_1$, $m_2$, $m_3$, $m_4$, and
$m_5$, such that the following equation for the Lorentz covariant 
scalar ${\widehat  {\cal G}} [  ({\widehat {\cal R}}) , ({\widehat {\cal 
R}}^{\prime}) ]$ takes the explicit form
\be
{\widehat  {\cal G}} [  ({\widehat {\cal R}}) , ({\widehat {\cal R}}^{\prime}) ] 
~=~ \left[\begin{array}{ccc}
~1 & ~0 &  ~0 \\
~0 & ~~1 &  -\, \fracm 13 \\
~0 &  -\, \fracm 13 &  ~~ 1  \\
\end{array}\right]      ~~~.
\label{Gdgt3}
\ee
to agree with adinka-based results in \cite{KIAS,KIAS2}. There exists an 
infinite number of such solutions. One solution is given by $m_1$ = $- \fracm 
1{\, 768 \,} $,  $m_2$ = $m_4$ =  $ \fracm 1{\, 1,536 \,}$ , and $m_3$ =
 $m_5$ =  0, but as
long as the following three equations are satisfied
\be \eqalign{
m_1 ~+~ 16 \, &m_3 ~+~ m_5 ~=~ - \fracm 1{\, 768 \,}   ~~~~, \cr
m_1 ~-~ 48 \, m_3 ~&+~ 8 \, m_4 ~+~  m_5 ~=~ \fracm 3{\, 768 \, }  ~~~, \cr
m_1 ~+~ 4 \, m_2 ~+~ 48 \, &m_3 ~-~ 4 \, m_4 ~-~ 3\, m_5 ~=~ - \fracm 1{\, 768
\,}
~~~,}
\label{CnStnts}
\ee
the result in (\ref{Gdgt3}) will be valid.

\section{The Gadget in Valise-Adinkra Networks} 
$~~~$
To readers who have been following our explorations for some time
\cite{KIAS, KIAS2,adnkholor,KIAS3}, the matrix denoted by $ {\widehat {\cal
G}} [ \, ({\widehat {\cal R}})  ,  ({\widehat {\cal R}}^{\prime})\, ] $ in (\ref{Gdgt3}) 
should be very familiar.  However, the initial appearance of this matrix (in 
the first of these references) was derived by methods that begin with valise 
adinkra networks.  It is useful to recount that derivation.

In the works of \cite{adnk1}-\cite{Top&3}, among others, it has been argued 
there exist networks which can be drawn in the forms of graphs that encode 
the representation theory of off-shell supersymmetrical multiplets in higher 
dimensional theories.  All such graphs to date have been constructed from 
a starting point of ones with cubical topology, but in order to irreducibly 
describe supersymmetry representations the nodes in such cubical 
graphs must be identified in a manner that uses error-correcting codes
\cite{adnkcodes}.  When the nodes of such graphs only appear at two 
levels, the graph is called a valise.  Three example (the so-called 
(C)-chiral, (V)-vector, and (T)-tensor) valise adinkras are shown below.
\vskip.2in  
$$
\vCent
{\setlength{\unitlength}{1mm}
\begin{picture}(-20,-140)
\put(-30,-16){\includegraphics[width=2.4in]{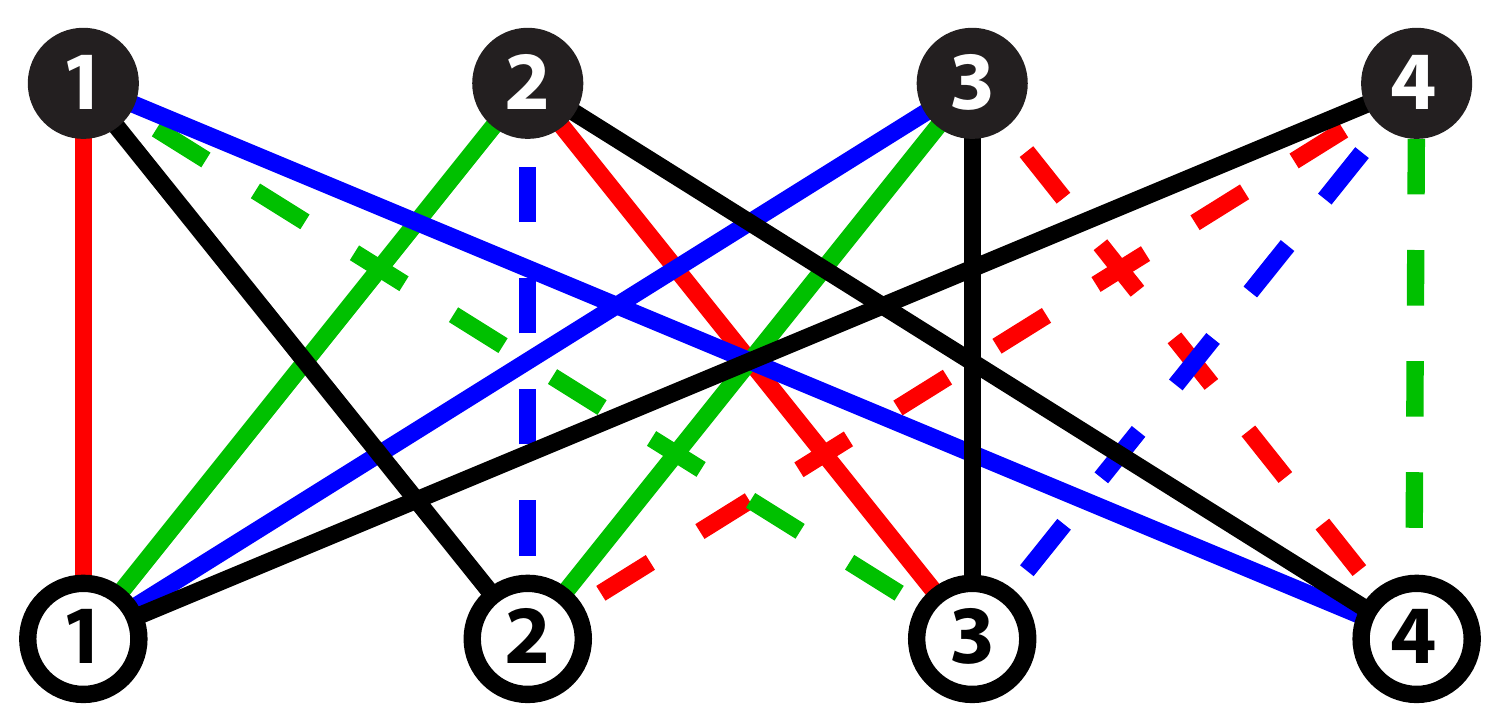}}
\put(-70,-24){\bf {Fig. \# 1: Illustration of a (C)hiral Valise Adinkra 
Network}}
\end{picture}}
$$
$$
\vCent
{\setlength{\unitlength}{1mm}
\begin{picture}(-20,-140)
\put(-30,-50){\includegraphics[width=2.4in]{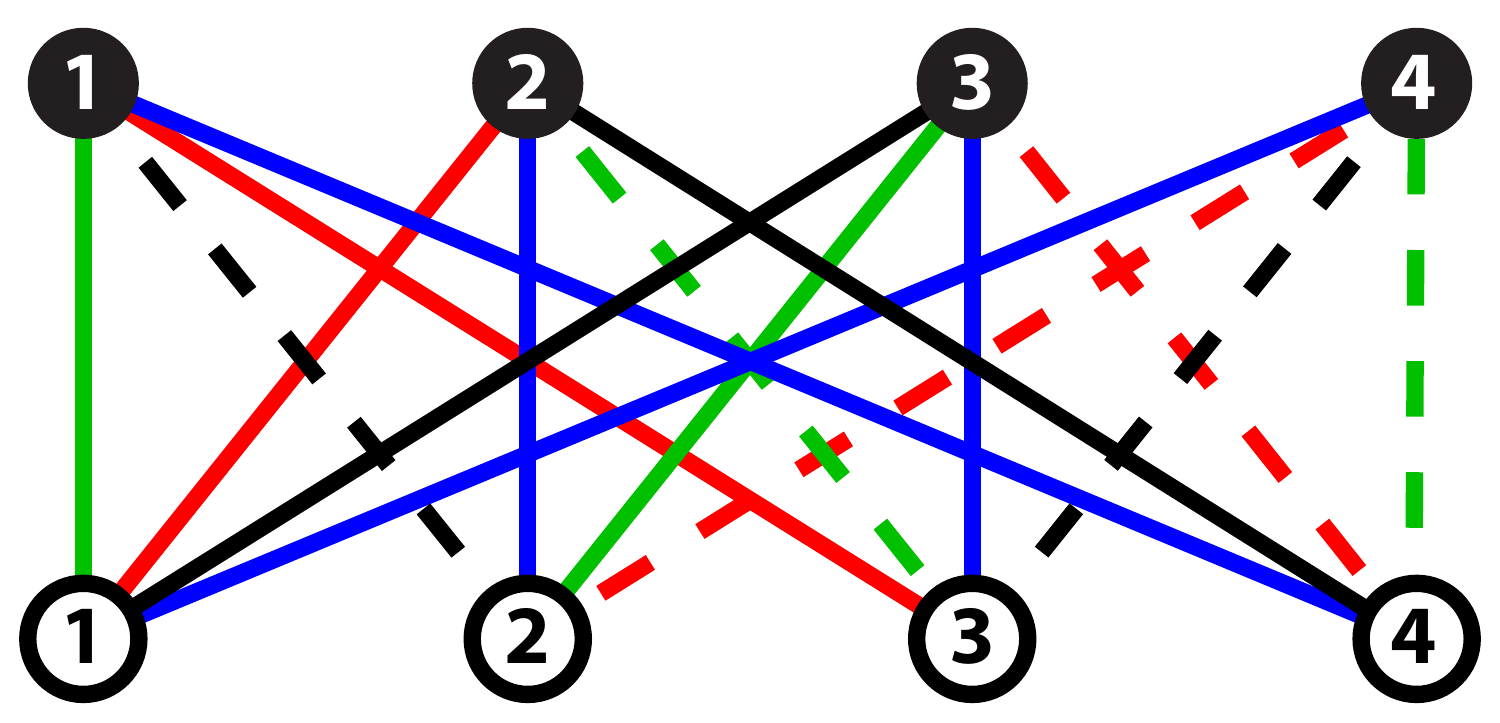}}
\put(-70,-58){\bf {Fig. \# 2: Illustration of a (V)ector Valise Adinkra 
Network}}
\end{picture}}
$$
$$
\vCent
{\setlength{\unitlength}{1mm}
\begin{picture}(-20,-140)
\put(-30,-84){\includegraphics[width=2.4in]{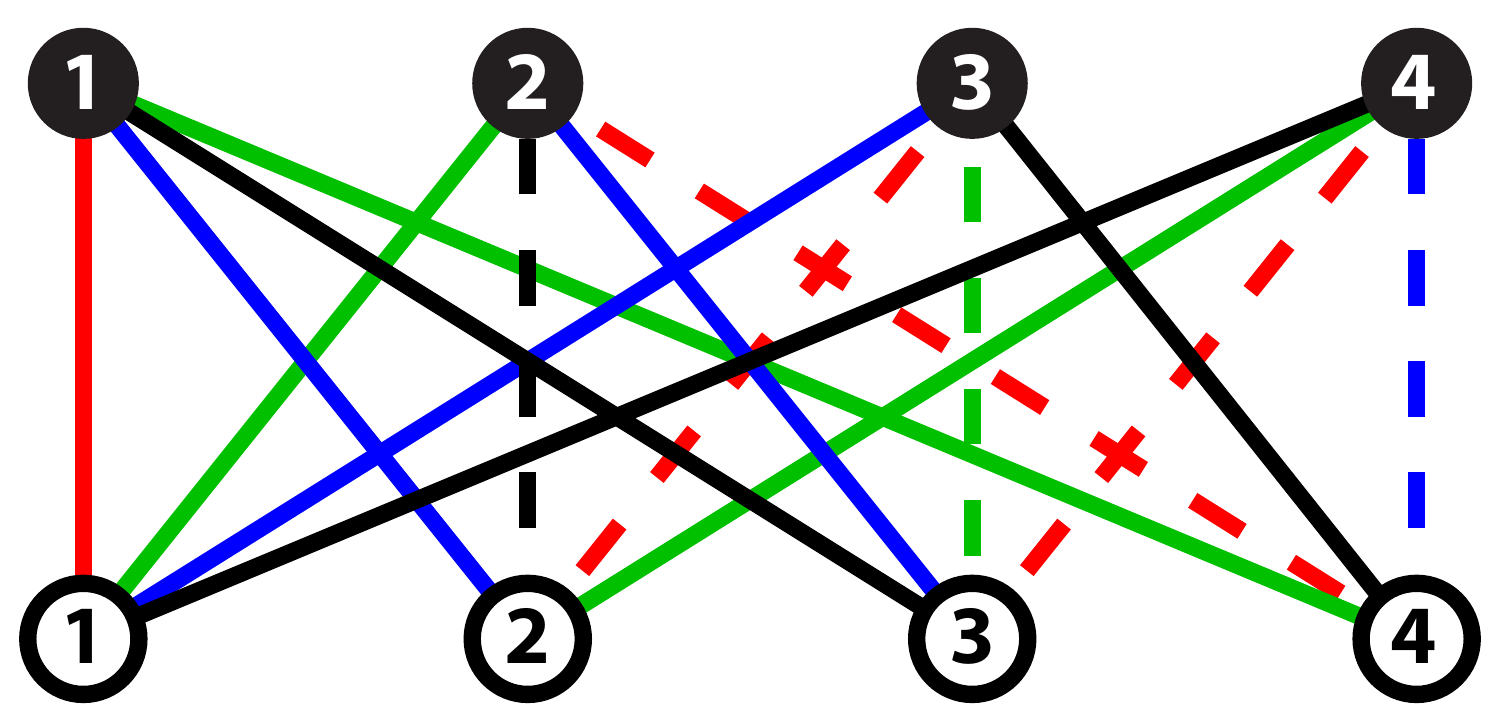}}
\put(-70,-92){\bf {Fig. \# 3: Illustration of a (T)ensor Valise Adinkra 
Network}}
\end{picture}}
$$
\vskip3.6in
A standard concept in graph theory is that a network possesses an ``adjacency
matrix.''  In our work, we have taken this concept a step further.  The links in adinkras
fall into equivalence classes.  Different classes are denoted by distinct colors in
the diagrams.  The distinct heights in the graphs denote distinct engineering dimensions
associated with the nodes.  Finally, links appeared dashed or not to indicate the 
presences of minus signs or not.  Thus, we replace the traditional adjacency 
graphs by more elaborate ``L-matrices'' and ``R-matrices.''  Alternately, if all
colored links are replaced by black lines and all dashing is dropped we recover a
standard adjacency matrix.  

For minimal representations, the R-matrices correspond to the matrix transposed 
version of the L-matrices.  For the respective three illustrations shown above, we 
have the following three respective sets of L-matrices shown below.  From here 
on we use a representation label $\cal R$ (without the `hat') to denoted different 
adrinkra network representations.  This is in distinction with the representation 
label $\widehat {\cal R}$ used for different supermultiplets in the last chapter.
From the work in \cite{KIAS,KIAS2} we have

\noindent
(C)~\hrulefill
$$  {
\left( {\rm L}{}_{1}\right) {}_{i \, {\hat k}}   ~=~
\left[\begin{array}{cccc}
~1 & ~0 &  ~~0  &  ~0 \\
~0 & ~0 &  ~~0  &  - 1 \\
~0 & ~1 &  ~~0  &  ~0 \\
~0 & ~0 &  ~- 1  &  ~0 \\
\end{array}\right] ~~~,~~~
\left( {\rm L}{}_{2}\right) {}_{i \, {\hat k}}   ~=~
\left[\begin{array}{cccc}
~0 & ~1 &  ~~0  &  ~  0 \\
~0 & ~ 0 &  ~~1  &  ~0 \\
-\, 1 & ~ 0 &  ~~0  &  ~0 \\
~ 0 & ~0 &  ~~0  &   - 1 \\
\end{array}\right]  ~~~, ~~~}
$$
\be  {
\left( {\rm L}{}_{3}\right) {}_{i \, {\hat k}}   ~=~
\left[\begin{array}{cccc}
~0 & ~0 &  ~~1  &  ~~0 \\
~0 & - 1 &  ~~0  &  ~~0 \\
~0 & ~0 &  ~~0  &  -\, 1 \\
~1 & ~0 &  ~~0  &  ~~0 \\
\end{array}\right] ~~~,~~~
\left( {\rm L}{}_{4}\right) {}_{i \, {\hat k}}   ~=~
\left[\begin{array}{cccc}
~0 & ~~0 &  ~~0  &  ~ \, \, 1 \\
~1 & ~~ 0 &  ~~0  &  ~~0 \\
~0 & ~~ 0 &  ~~1  &  ~~0 \\
~ 0 & ~~~1 &  ~~0  &   ~~0  \\
\end{array}\right]  ~~.  }
 \label{chiD0F}
\ee
\noindent
(V)~\hrulefill
$$ {
\left( {\rm L}{}_{1}\right) {}_{i \, {\hat k}}   ~=~
\left[\begin{array}{cccc}
~0 & ~1 &  ~ 0  &  ~ 0 \\
~0 & ~0 &  ~0  &  -\,1 \\
~1 & ~0 &  ~ 0  &  ~0 \\
~0 & ~0 &  -\, 1  &  ~0 \\
\end{array}\right] ~~~,~~~
\left( {\rm L}{}_{2}\right) {}_{i \, {\hat k}}   ~=~
\left[\begin{array}{cccc}
~1 & ~ 0 &  ~0  &  ~ 0 \\
~0 & ~ 0 &  ~1  &  ~ 0 \\
 ~0 & - \, 1 &  ~0  &   ~ 0 \\
~0 & ~0 &  ~0  &  -\, 1 \\
\end{array}\right]  ~~~, }
$$
\be {~~~~} {
\left( {\rm L}{}_{3}\right) {}_{i \, {\hat k}}   ~=~
\left[\begin{array}{cccc}
~0 & ~0 &  ~ 0  &  ~ 1 \\
~0 & ~1 &  ~0  &   ~0 \\
~0 & ~0 &  ~ 1  &  ~0 \\
~1 & ~0 &  ~0  &  ~0 \\
\end{array}\right] ~~~~~~,~~~
\left( {\rm L}{}_{4}\right) {}_{i \, {\hat k}}   ~=~
\left[\begin{array}{cccc}
~0 & ~0 &  ~1  &  ~ 0 \\
-\,1 & ~ 0 &  ~0  &  ~ 0 \\
 ~0 & ~0 &  ~0  &   - \, 1 \\
~0 & ~1 &  ~0  &  ~  0 \\
\end{array}\right]  ~~~, }
\label{V1D0E}
\ee
\noindent
(T)~\hrulefill
$$  {
\left( {\rm L}{}_{1}\right) {}_{i \, {\hat k}}   ~=~
\left[\begin{array}{cccc}
~1 & ~0 &  ~0  &  ~0 \\
~0 & ~0 &  -\, 1  &  ~ 0 \\
~0 & ~0 &  ~0  &  -\,1 \\
~0 & -\,1 &  ~ 0  &  ~0 \\
\end{array}\right] ~~~,~~~
\left( {\rm L}{}_{2}\right) {}_{i \, {\hat k}}   ~=~
\left[\begin{array}{cccc}
~0 & ~1 &  ~0  &  ~  0 \\
~0 & ~ 0 &  ~0  &  ~ 1 \\
~0 & ~ 0 &  -\,1  &  ~ 0 \\
 ~ 1 & ~0 &  ~0  &   ~ 0 \\
\end{array}\right]  ~~~,
}    $$
\be {~~~~} {
\left( {\rm L}{}_{3}\right) {}_{i \, {\hat k}}   ~=~
\left[\begin{array}{cccc}
~0 & ~0 &  ~1  &  ~0 \\
~1 & ~0 &  ~ 0  &  ~ 0 \\
~0 & ~1 &  ~0  &   ~0 \\
~0 & ~0 &  ~ 0  &  -\, 1 \\
\end{array}\right] ~~~~~~,~~~
\left( {\rm L}{}_{4}\right) {}_{i \, {\hat k}}   ~=~
\left[\begin{array}{cccc}
~0 & ~0 &  ~0  &  ~  1 \\
~0 & -\, 1 &  ~0  &  ~ 0 \\
~1 & ~ 0 &  ~0  &  ~ 0 \\
 ~0 & ~0 &  ~1  &   ~ 0 \\
\end{array}\right]  ~~~,  }
\label{tenD0F}
\ee
and given the L-matrices associated with any of the adinkras, we can define
another set of matrices (the $\Tilde V$ matrices \cite{KIAS2}) via the equations.
\be {
\eqalign{
({\Tilde V}{}_{\rI \, \rJ}^{({\cal R})} ){}_\hi{}^\hk ~=~ \fracm 12 \, \left[ (\,{\rm R
}^{({\cal R})}_\rI\,)_\hi{}^j\>(\, {\rm L}^{({\cal R})}_\rJ\,)_j{}^\hk - (\,{\rm R}^{(
{\cal R})}_\rJ\,)_\hi{}^j\>(\,{\rm L}^{({\cal R})}_\rI\,)_j{}^\hk  \, \right] ~~~.
}} \label{VmtRX1}
\ee
Since the L-matrices are dependent on which adinkra representation is
taken as a starting point, this is also the case for the $\Tilde V$ matrices
and we indicate this by including a representation label ${\cal R}$.
Explicit calculations for each set leads to \cite{KIAS,KIAS2}
\be {
\eqalign{
 ({\Tilde V}{}_{\rI \, \rJ}^{({\cal R})} ){}_\hi{}^\hk
  &= i\, \Big[ \,\ell^{({{\cal R}})1}_{\rI\rJ}\, (\a^1){}_{\hi}{}^\hk
   +  \ell^{({{\cal R}})2}_{\rI\rJ}\, (\a^2){}_{\hi}{}^\hk 
   ~+~  \ell^{({{\cal R}})3}_{\rI\rJ}\, (\a^3){}_{\hi}{}^\hk   \, \Big] 
  \cr
 &~~~+~  i\, \Big[ \, {{\Tilde \ell}^{({\cal R})}}_{
 \rI\rJ}{}^{1}\, (\b^1){}_{\hi}{}^\hk
 ~+~
  {{\Tilde \ell}^{({\cal R})}}_{\rI\rJ}{}^{2}\, 
 (\b^2){}_{\hi}{}^\hk  \,+\,  {{\Tilde \ell}^{({\cal R})}}_{\rI\rJ}{}^{3}\, 
 (\b^3){}_{\hi}{}^\hk 
  \, \Big]  ~~.
} } \label{VmtRX2}
\ee
with the non-vanishing entries for each representation taking the forms
 \be {
\begin{array}{cccccc}
 \ell _{12}^{(C)1}=1 & \ell _{13}^{(C)2}=1 & \ell _{14}^{(C)3}=1 & \ell _{23}^{(
 C)1}=1 & \ell _{24}^{(C)2}=-1 & \ell_{34}^{(C)3}=1  ~~~~, \\
 \Tilde{\ell }_{12}^{(V)1}=-1 & \Tilde{\ell }_{13}^{(V)2}=1 & \Tilde{\ell }_{14}^{(
 VM)3}=-1 & \Tilde{\ell }_{23}^{(V)1}=1 &
 \Tilde{\ell }_{24}^{(V)2}=1 & \Tilde{\ell }_{34}^{(V)3}=1  ~~~~, \\
 \Tilde{\ell }_{12}^{(T)1}=1 & \Tilde{\ell }_{13}^{(T)2}=1 & \Tilde{\ell
 }_{14}^{(T)3}=1 & \Tilde{\ell }_{23}^{(T)1}=-1 &
\Tilde{\ell }_{24}^{(T)2}=1 & \Tilde{\ell }_{34}^{(T)3}=-1 ~~.
\end{array}  }
\label{els}
\ee
The 4 $\times$ 4 matrices ${{\bm \a}}^1$, ${{\bm \a}}^2$, ${{\bm \a}}^3$, 
${{\bm \b}}^1$, ${{\bm \b}}^2$, ${{\bm \b}}^3$ which appear in (\ref{VmtRX2}) 
can be written as
\be \eqalign{
&~~~~{{\bm \a}}^1 =~ {\bm \s}^2 \otimes {\bm \s}^1 ~~~\,~~,~~~~~ {{\bm \a}}^2 = 
{\bf I}{}_2  \otimes {\bm \s}^2  ~~   ~~~~\,~~,~~~~{{\bm \a}}^3 = {\bm \s}^2 
\otimes {\bm \s}^3 ~~~~~~~~~, \cr
 &~~~~{{\bm \b}}^1 =~ {\bm \s}^1 \otimes {\bm \s}^2 ~~~\,~~,~~~~~ {{\bm \b}}^2 = 
 {\bm \s}^2 \otimes {\bf I}{}_2   ~~   ~~~~~~~,~~\,~{{\bm \b}}^3 = {\bm \s}^3 \otimes 
 {\bm \s}^2~~~~~~\,~~~.   
}  \label{Defs}
\ee
where the outer product conventions are in \cite{G-1}.  These matrices
satisfy the identities
\be  \eqalign{ {~~~~~~~}
{\bm \a}^{\Hat I} \, {\bm \a}^{\Hat K} ~&=~ \delta{}^{{\Hat I} \, {\Hat K}} \, {\bm {\rm I}}{}_4
~+~ i \, \epsilon{}^{{\Hat I}  \, {\Hat K} \, {\Hat L}} \, {\bm \a}^{\Hat K} ~~~,~~~ {\bm \b}^{\Hat 
I} \, {\bm \b}^{\Hat K} ~=~ \delta{}^{{\Hat I} \, {\Hat K}} \,  {\bm {\rm I}}{}_4 ~+~ i \, 
\epsilon{}^{{\Hat I}  \, {\Hat K} \, {\Hat L}} \, {\bm \b}^{\Hat K}  ~~,   \cr
& {~} {\rm {Tr}} \big( \,  {\bm \a}^{\Hat I}  \, {\bm \a}^{\Hat J}  \, \big) ~=~ {\rm {Tr}} \big( \, 
{\bm \b}^{\Hat I}  \,  {\bm \b}^{\Hat J} \, \big) ~=~ 4\,  \delta{}^{{\Hat I} \, {\Hat J}} ~~,~~ 
\ {\rm {Tr}} \big( \,  {\bm \a}^{\Hat I}  \, {\bm \b}^{\Hat J}  \, \big) 
~=~ 0 ~~,   \cr
& {~~~~~~~~~~~~~~~~~~~~~} {\rm {Tr}} \big( \,  {\bm \a}^{\Hat I}   \, \big) ~=~ {\rm {Tr}} \big( \, 
{\bm \b}^{\Hat I}  \,  \, \big) ~=~ 0 ~~~.
}   \label{Trcs}
\ee

Finally, a gadget for the valise adinkra network can be defined by
\be 
{\cal G}\left[  ({\cal R}) , \,  ({\cal R}^{\prime}) \right] \,=~- \, \fracm 1{48} \,
\sum_{{\rm I}, {\rm J}} {\rm {Tr}} \left[   {\Tilde V}{}_{\rI \, \rJ}^{({\cal R})}
 {\Tilde V}{}_{\rI \, \rJ}^{({\cal R}^{\prime})} \right] \,=\,  \fracm 1{12} \,
\sum_{{\rm I}, {\rm J},  {\hat a} } \, \left[ ~ {\ell}_{{\rm I}{\rm J}}^{\,
({\cal R}) \hat{a}} \,  {\ell}_{{\rm I}{\rm J}}^{\, ({\cal R}^{\prime})
\hat{a}}  ~+~   \Tilde{\ell}_{{\rm I}{\rm J}}^{\, ({\cal R}) \hat{a}}
\,  \Tilde{\ell}_{{\rm I}{\rm J}}^{\, ({\cal R}^{\prime}) \hat{a}}  ~ \right]    ~~~,
 \label{Gdgt4}
\ee
and upon using the information in (\ref{VmtRX2}), we find
\be 
{\cal G}\left[  ({\cal R}) , \,  ({\cal R}^{\prime}) \right] 
~=~ \left[\begin{array}{ccc}
~1 & ~0 &  ~0 \\
~0 & ~~1 &  - \, \fracm13 \\
~0 & - \, \fracm13 &  ~~ 1  \\
\end{array}\right]  
~~~.
 \label{Gdgt5}
\ee

\section{Why The Commutator of Supercharges?} 
$~~~$
In an epochal paper, D.~Gross and R.~Jackiw \cite{G-J} noted a particular 
mathematical quantity in the representation theory of Lie Algebras, plays a 
prominent role with regard to anomalies in gauge theories.  The quantity 
in question can be called the ``d-coefficients tensor'' (following conventions
that arise in the context of the su(3) Lie algebra).  For the purposes of our 
discussion we will write this in the form 
as
\be
d{}_{A \, B \, C}^{({\cal R})}  ~=~ \fracm 12 \, {\rm {Tr}} \left[ \,   \{ \,  {\bm t}{}_{A
}^{({\cal R})} ~,~  {\bm t}{}_{B}^{({\cal R})}  \, \}  \,   {\bm t}{}_{C}^{({\cal R})}  \,   
\right] ~~~,  \label{eQ1}
\ee
where the notation is indicative of several relevant features.  In this expression
there appear some matrices ${\bm t}{}_{A}^{({\cal R})}$ (with $A$ = 1, 2, $\dots$
$p$ for some integer $p$) in a representation $\cal R$ of some Lie algebra.  
This way of defining the d-coefficients has the advantage that for any set of
matrices ${\bm t}{}_{A}^{({\cal R})}$, this provides a a well-defined way to 
explicitly calculate them.  We may let the symbol $d_{({\cal R})}$ denote the 
dimension of $({\cal R})$ and $d_{({\cal R}^{\prime})}$ denote the dimension 
of $({\cal R}^{\prime})$.  Another result that follows from this definition,
can be seen from the following consideration.  

If the matrices $ {\bm t}{}_{A}^{({\cal R}^{\prime})} $ of one representation 
(denoted by $({\cal R}^{\prime})$) are related to the matrices $ {\bm t}{}_{A
}^{({\cal R})} $ via equations of the form
\be
{\bm t}{}_{A}^{({\cal R}^{\prime})} ~=~ {\cal S}{}^{-1} \, {\bm t}{}_{A}^{({\cal R})}
 {\cal S} \, ~~~,
  \label{eQ2}
\ee
for some matrix $ {\cal S}$ and its inverse, then 
\be
d{}_{A \, B \, C}^{({\cal R}^{\prime})}  ~=~
d{}_{A \, B \, C}^{({\cal R})}   ~~~.
 \label{eQ3} \ee
However, the ``d-coefficients tensor'' can be calculated for any representation
and the representation $({\cal R}^{\prime})$ need not be restricted to satisfy (\ref{eQ2}).

Clearly the values of these coefficients depend on the choice of
how one orders the ${\bm t}$-matrices and to utilize a quantity that does not
depend on the ordering we define a mathematical ``gadget'' (denoted by 
${\tilde g} ({\cal R},\, {\cal R}^{\prime})$) on these representation spaces via 
the equation
\be
{\tilde g} ({ {\cal R}},\, { {\cal R}}^{\prime}) ~=~ {\cal N}_0 \sum_{A, \, B, \, C}
 d{}^{({\cal R})} _{A \, B} {}^{C} \,  d{}^{({\cal R}^{\prime})} _{A \, B} {}^{C}
 ~~~,
  \label{eQ4}
\ee
where ${\cal N}_0$ is a normalization constant whose value is fixed
by requiring that ${\tilde g} ({\cal R},\, {\cal R})$ = 1 when $({\cal R})$
denotes a minimal irreducible representation.  If the d-coefficients 
defined in (\ref{eQ1}) are real, the gadget assigns a real number to 
the pair of representations $({\cal R})$ and $({\cal R}^{\prime})$.  
Furthermore, whenever $({\cal R})$ = $({\cal R}^{\prime})$, the gadget 
assigns a non-negative number if the d-coefficients are real.

A simple example of this formalism can be seen in the case where
the Lie algebra is su(3), with $({\cal R})$ = $\{ 3\}$, and $({\cal R}^{\prime})$ 
= $\{ {\Bar 3}\}$.  The form of the gadget in this case is given by a 
2 $\times$ 2 matrix of the form
\be
{\tilde g} ({{\cal R}},\, { {\cal R}}^{\prime})
~=~
\left[\begin{array}{cc}
1 & - 1 \\
- 1 & ~ 1 \\
\end{array}\right] ~~~.
 \label{eQ6}
\ee
Here the quantities ${ {\cal R}}$ and ${ {\cal R}}^{\prime}$
are to be regarded as indices that each take on the values $\{ 3 \}$
and $\{ {\Bar 3}\}$.

This example shows that when  $({\cal R})$ $\ne$ $({\cal R}^{\prime})$,
the gadget can produce real but negative values.  Finally, although we
will not carry out the calculations, a further informative example consists
of working out explicitly the values of the gadget for the case of $({\cal 
R})$ = $\{ 3\}$, and $({\cal R}^{\prime})$ = $\{ {8}\}$. 

When there exists two representations $({\cal R})$ $\ne$ $({\cal R}^{\prime})$
of a Lie algebra, where both are represented by $d$ $\times$ $d$ matrices,
in addition to forming the traditional d-coefficients, there is another possibility
to form a rank four tensor
\be
{\cal H}{}_{A \, B \, C \, D}^{[(  {\cal R}),({\cal R}^{\prime})]} ~=~ {\rm {Tr}} 
\left[ \,   \{ \,  {\bm t}{}_{A}^{({\cal R})} ~,~  {\bm t}{}_{B}^{({\cal R})}  \, \} \, \{ 
\,  {\bm t}{}_{C}^{({\cal R}^{\prime})} ~,~  {\bm t}{}_{D}^{({\cal R}^{\prime})} 
\, \} \right] 
~~~.  \label{eQ7}
\ee

In the works of  \cite{KIAS, KIAS2,adnkholor,KIAS3}, the concept of the gadget 
was extended beyond matrix representations of compact Lie algebras to the 
realm of adinkra network valise graphs and 0-brane reduced four 
dimensional minimal SUSY supermultiplets.  These works were enabled
due to the elucidation of a rank four tensor that exists in these systems
which was used to play the role of the d-coefficients.  This rank four tensor 
was given the name of the ``holoraumy'' tensor and it is analogous to the
Lie algebraic tensor defined in (\ref{eQ7}).

\section{Observations and Summary} 
$~~~$ 
To recapitulate, we have shown the existence of a metric\footnote{This is one 
member of a class of such metrices.} over the 
representation space of minimal off-shell 4D, $\cal N$ = 1 supermultiplets, given by
\be
 \eqalign{ {~~}
{\widehat  {\cal G}} [  ({\widehat {\cal R}}) , ({\widehat {\cal R}}^{\prime}) ] 
~&=~ - \fracm 1{768} {\bm{\Large \{ }}\, \,  [{ {\bm H}}{}^{\mu}{}^{({\widehat 
{\cal R}})} ]{}_{a \, b \, c}{}^d  \, [ { {\bm H}}{}_{\mu}{}^{({\widehat {\cal 
R}}^{\prime})} ]{}^{a \, b}{}_{d}{}^c   \cr
&~~~~~~~~~~~~~~~~-~ \fracm 12 \,   (\gamma^{\a})_c^{~e}  \,  [{ {\bm H}}{}^{\mu}{
}^{({\widehat {\cal R}})} ]{}_{a \, b \, e}{}^f  \, (\gamma_{\a})_f^{
~d} \, [ { {\bm H}}{}_{\mu}{}^{({\widehat {\cal R}}^{\prime})} ]{}^{a \, 
b}{}_{d}{}^c        \cr
&~~~~~~~~~~~~~~~~-~ \fracm 12 \,    (\gamma^5 \gamma^{\a})_c^{~e}  \,  [{ {\bm 
H}}{}^{\mu}{}^{({\widehat {\cal R}})} ]{}_{a \, b \, e}{}^f  \, (\gamma^5
\gamma_{\a})_f^{~d} \, [ { {\bm H}}{}_{\mu}{}^{({\widehat {\cal R}}^{\prime
})} ]{}^{a \, b}{}_{d}{}^c   \,  {\bm{\Large \} }} ~~~,
} 
\ee
(where the coefficients are defined by (\ref{HCS}), (\ref{HVS}), and (\ref{HTS}))
has elements identical (over the different supermultiplet representations) to those
in the metric
\be  \eqalign{
{\cal G}\left[  ({\cal R}) , \,  ({\cal R}^{\prime}) \right] ~&=~ - \,  \fracm 1{48} \, 
\sum_{{\rm I}, {\rm J} } \, {\rm {Tr}} \left[ \, {\Tilde V}{}_{\rI \, \rJ}^{({\cal R})}  
{\Tilde V}{}_{\rI \, \rJ}^{({\cal R}^{\prime})}   \, \right]  ~~~,
} \ee
(where these coefficients are defined by (\ref{VmtRX1}), (\ref{VmtRX2}), and 
(\ref{els})) for the adjacency matrices of three corresponding adinkra 
networks shown in illustrations (\# 1), (\# 2), and (\# 3). 

We have thus, {\em {for the first time and directly in four dimensions}}, realized
the possibility to define a consistent geometrical viewpoint of the three minimal 
off-shell $\cal N$ = 1 supermultiplets as elements in a representation space 
with a metric.

We established an equation
\be
 {\widehat {\cal G}}[  ({\widehat {\cal R}})  ,  (
 {\widehat {\cal R}}^{\prime})\, ]
~=~
 { {\cal G}} \left[ ({\cal R}) , ({\cal R}^{\prime}) \, \right]
 ~~~~, \label{Greps}
\ee
which provides a realization of the concept of ``SUSY holography,''
(i.e. the proposal that adinkras are holograms of supermultiplets).  The
existence of such an equation is critical for the entire program we initiated
with the work of \cite{ENUF7}.

The emergence of a four dimensional holoraumy structure among 
the minimal off-shell $\cal N$ = 1 supermultiplets can be seen from the Lorentz 
representation structure of the three equations in (\ref{HCSf}), (\ref{HVSf}), and 
(\ref{HTSf}).  Due to the appearance of the commutator of the SUSY charges 
on the right hand side of each equation, one should generally expect the left 
hand side of the equations to contain the scalar, psuedoscalar, and axial vector 
Lorentz representations induced by the $a$ and $b$ indices.  It is striking
that {\em {only}} the axial vector occurs for the chiral multiplet fermion, while the
calculations for the fermions in the other two supermultiplets contains {\em {all}}
{\em {three}} expected representations.

We have fixed our normalizations so ${\widehat {\cal G}}$ implies the 
CS, VS, and TS representations correspond to unit vectors in a representation
space.  The unit vector representing the CS representation is orthogonal to the 
unit vectors representing the VS, and TS representations.  We define an angle 
between any two of the 4D, $\cal N$ = 1 supermultiplet representations $({\widehat 
{\cal R}})$ and $({\widehat {\cal R}}^{\prime})$ 
via the definition
\be {
cos \left\{ \theta [({\widehat {\cal R}})\,  , \, ({\widehat {\cal R}}^{\prime})] \right\} ~=~
{{{\widehat {\cal G}} [ \, ({\widehat {\cal R}}) , ({\widehat {\cal R}}^{\prime}) \,] } 
\over {~ {\sqrt{{\widehat {\cal G}} [ \, ({\widehat {\cal R}}) , ({\widehat {\cal R}}) 
\, ]}} \, {\sqrt{ {\widehat {\cal G}} [ \, ({\widehat {\cal R}}^{\prime}) , ({\widehat {\cal 
R}}^{\prime}) ]}}~~ } } ~~~.  }   \label{M4}
\ee
The angles between the VS, and TS representations can be read from the matrix 
given in (\ref{Gdgt3}) or (\ref{Gdgt5}) to have the common of $\theta_{TV}$ where
\be
cos (\theta_{TV}) ~=~ -  \, \fracm 13  ~~~~.
\label{VTthetan}
\ee
On the space of the CS, VS, and TS representations, this is illustrated below
in agreement with the works of \cite{KIAS,KIAS2}. 

The process of removing degrees of freedom by dimensional reduction 
is an example of an `injection.''  By this we mean there exist many consistent
prescriptions for constructing different adinkra shadows for any single higher 
dimensional supermulti-

\newpage
$$
\vCent
{\setlength{\unitlength}{1mm}
\begin{picture}(-20,-140)
\put(-45,-66){\includegraphics[width=3.8in]{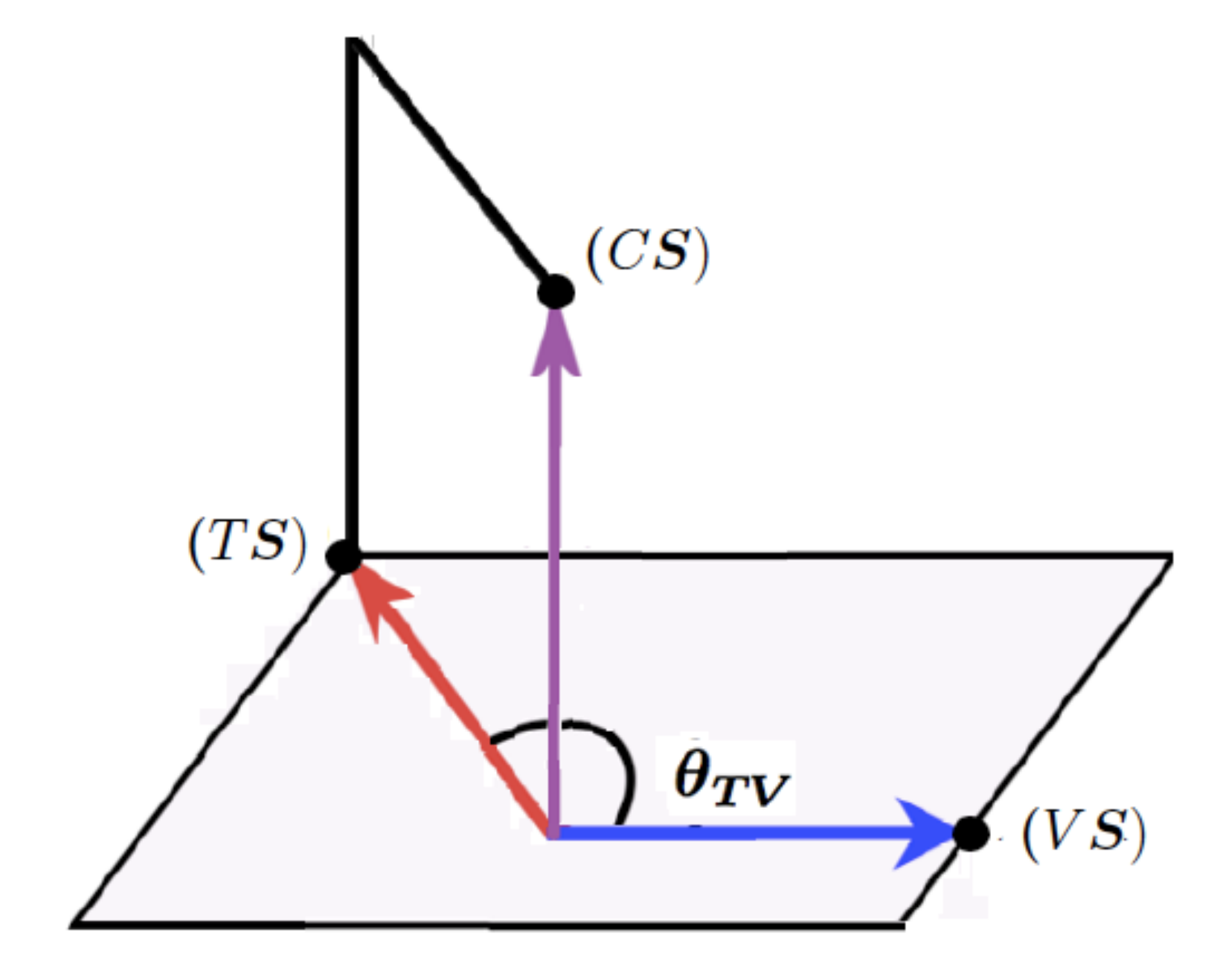}}
\put(-74,-70){\bf {Fig. \# 4: Illustration of the CS-VS-TS subspace using the 
$\widehat {\cal G}$ metric}}
\end{picture}}
$$
\vskip2.6in \noindent
plet.  In terms of thinking of the process as a map, there exist many consistent 
processes that take one supermultiplet and inject it into a ``sea'' of adinkra 
networks.  Explicit examples of this can now be completely and thoroughly 
discussed due to the results uncovered in the work of \cite{permutadnk}.

The key to our progress of this paper is the existence of the four dimensional 
Lorentz covariant holoraumy tensor dependent upon four integers p, q, r, and 
s in the following formula
\be  \eqalign{ {~~~~}
\left[{ {\bm H}}{}^{\mu}( {\rm p},  {\rm q}, {\rm r},   {\rm s}  \,) \right]{}_{a \, b \, 
c}{}^d  ~&=~ -i 2\, {\rm p} \, C_{ab} \, (\gamma^{\m})_c{}^{d}  ~-~ i 2 \,  {\rm q}
 \, (\gamma^{5})_{ab} (\gamma^{5}\gamma^{\m})_{c}{}^{d}     \cr
 &~~~~~  -i \,  {\rm r} \, (\gamma^5 \gamma^{\nu} )_{ab} \, (\gamma^5 \, [
 \gamma {}_{\nu}\,, \, \gamma^{\mu} ])_{c}{}^{d}   ~-\, i \, 2 \, {\rm s} \,  (\gamma^5 
 \gamma^{\mu} )_{ab} \, (\gamma^5)_{c}{}^{d}   ~~~,
}   \label{4DHs}
\ee
and corresponding to the three vectors illustrated above we have
\begin{center}
\renewcommand\arraystretch{1.2}
\begin{tabular}{|c|c|c|c|c| }\hline
${\rm {Supermultiplet}}$  & ${\rm {p}}$  & ${\rm {q}}$ & ${\rm {r}}$  & ${\rm {s}}$ 
\\ \hline
\hline
${\rm {CS}}$ & $0$ &  $0$  &  $+\, 1$ &  $ 0$  \\ \hline
${\rm {VS}}$ & $+\, 1$ &  $+\, 1$  &  $0$ &  $ + \, 1$  \\ \hline
${\rm {TS}}$ & $-\, 1$ &  $+\, 1$  &  $0$ &  $ - \, 1$  \\ \hline
\end{tabular}
\end{center}
\vskip.1in
\centerline{{\bf Table I}}
that encodes the distinct supermultiplets in an extremely compact
manner. 

This method of classifying the minimal off-shell supermultiplets makes a finer
distinction than we gave in the works of \cite{G-1} or \cite{adnkColor}.
In the first of these, a quantum number $\chi{}_{\rm o}$ (``Kye-Oh'') was
introduced initially via adinkra networks and
extended to supermultiplets in the second.  Comparing these previous
works to the formula in (\ref{4DHs}), we see
\be
\chi{}_{\rm o} ~=~  (- \, 1){}^{{\rm s}^2}   ~~~~,
\ee 
with regard to the minimal off-shell supermultiplet representations.

If we include the parity reflected and variant representation versions of the
minimal (see \cite{KIAS3}) supermultiplets there are only eight such 
representations.  It is thus possible to extend the minimal supermultiplet 
representation label $\widehat {\cal R}$ that appears on the left hand side 
of (\ref{Greps}) to cover an eight dimensional vector space.  On the other 
hand the work of  \cite{permutadnk} implies the existence of 1,536 sets 
of consistent adinkra networks based on the existence of a Coxeter group.  
This means it is possible to extend the minimal adinkra network representation 
label $\cal R$ that appears on the right hand side of (\ref{Greps}) to cover 
a 1,536 dimensional vector space!

We are now in position to state a conjecture about 4D, $\cal N$ = 1 minimal 
off-shell supermultiplets and 1d, $N$ = 4 minimal adinkras constructed from 
a Coxeter group.   Let ${\widehat {\cal R}}{}_1$, $\dots$, ${\widehat {\cal R}}
{}_8$ denote any of the minimal off-shell 4D, $\cal N$ = 1 supermultiplets.  
Let ${\cal R}_1$, $\dots$, ${\cal R}_{1,536}$ denote any of the minimal four 
color adinkra based on the Coxeter group described in \cite{permutadnk}.  
Let ${\cal R}{}_1^*$, $\cdots$, ${\cal R}{}_8^* $ denote any eight among the 
1,536 adinkra representations.

We conjecture that if the equation in (\ref{Greps}) is satisfied for some
ordering of the supermultiplets ${\widehat {\cal R}}{}_1$, $\dots$, 
${\widehat {\cal R}}{}_8$ together with an appropriate ordering of
adinkra network representations ${\cal R}{}_1^*$, $\cdots$, ${\cal R}{}_8^* $, 
then there must exist a projection
operator $\cal P$ such that
\be  \eqalign{
{\cal P}: {\widehat {\cal R}}{}_1 ~ &\to ~ {\cal R}  {}_1^*   ~~~~, \cr
{\cal P}: {\widehat {\cal R}}{}_2 ~ &\to ~ {\cal R}  {}_2^*   ~~~~, \cr
&\vdots  \cr
{\cal P}: {\widehat {\cal R}}{}_8 ~ &\to ~ {\cal R}  {}_8^*   ~~~~.
} \ee

An important implication of this conjecture is that the illustrations (\# 1), 
(\# 2), and (\# 3) provide only one consistent set of adinkra representations
of the minimal off-shell 4D, $\cal N$ = 1 supermultiplets.  Any three adinkra networks
that preserve the conditions in (\ref{Greps}) can act as the shadows 
for the four dimensional supermultiplets.

There can easily arise a problem associated with starting from a chosen
set of adinkra networks along with the data they contain and attempting to 
reconstruct the higher dimensional supermultiplets to be associated 
with these chosen
networks.  Only sets of adinkras that satisfy the equation in (\ref{Greps})
should be identified with the higher dimensional supermultiplets.  Stated
another way, the condition in (\ref{Greps}) acts as a filter for how one
begins from adinkras and then uses these as a tool to reconstruct higher
dimensional supermultiplets.

We believe ultimately, the condition in (\ref{Greps}) will play a very
important role in the program that has been initiated by the work of
\cite{adnkgeo} which has as its aim to place the representation theory
of supersymmetry into the context of algebraic geometry and Riemann
surfaces.

In this work, we have built upon the foundation that has been provided
by careful and detailed studies of the case of four dimensional $\cal N$
= 1 off-shell minimal supermultiplets and minimal four-color adinkras.
There is still more work to be done in order to extend the formula in
(\ref{4DHs}) for the Lorentz covariant holoraumy tensor to cover other 
four dimensional $\cal N$ = 1 off-shell supermultiplets...and beyond.  
It is our expectation that for more complication supermultiplets, the
4D holoraumy tensors we have found will likely have to by augmented
by additional ones, perhaps with a different Lorentz representation
structure.  However, with the success demonstrated in this current work, 
we are confident about the success of such efforts to be undertaken in 
the future.

 \vspace{0.2in}
 \begin{center}
\parbox{4in}{{\it ``All truths are easy to understand once they are \\ ${~}$ 
discovered; the point is to discover them.
 \\ ${~}$ 
\\ ${~}$ }\,\,-\,\, Galileo Galilei $~~~~~~~~~$}
 \parbox{4in}{
 $~~$}  
 \end{center}
$$~~$$

\noindent
{\bf Acknowledgements}\\[.1in] \indent
This work was partially supported by the National Science Foundation grant 
PHY-13515155.  This research was also supported in part the University of 
Maryland Center for String \& Particle Theory.  Additional acknowledgment 
is given by T.\ Grover, M.\ D.\ Miller-Dickson, B.\ A.\ Mondal, A.\ Oskoui, S.\ 
Regmi, E.\ Ross, and R.\ Shetty to the Center for String and Particle Theory 
(CSPT), as well as recognition for their participation in 2015 SSTPRS 
(Student Summer Theoretical Physics Research Session) program.  
Adinkras in this work were created by T.\ H{\"u}bsch.

\newpage
\noindent
{\bf{Appendix: Components of the 4D Gadget Calculations}}

\indent
In the appendix, we provide some of the intermediate calculations that lead to 
the results in
(\ref{CnStnts}).  We begin with the 4D holoraumy tensors defined by
Next, we need to perform a series of ``conjugation'' transformations on each
of these by acting with the matrices $(\gamma^{\a})$, $([
\gamma^{\a}  ~,~ \gamma^{\b}])$, $(\gamma^{5} \gamma^{\a})$, 
and $(\gamma^{5})$ respectively.
$$
 \eqalign{
(\gamma^{\a})_c^{~e}  \,  \left[{ {\bm H}}{}^{\mu}{}^{(CS)}\right]{}_{a \, b \, 
e}{}^f  \, (\gamma_{\a})_f^{~d} 
~&=~  0 ~~~~, \cr 
([\, 
\gamma^{\a}  ~,~ \gamma^{\b} \,]))_c^{~e}  \,  \left[{ {\bm H}}{}^{\mu}{}^{(CS)
}\right]{}_{a \, b \, e}{}^f   \, ([\, \gamma_{\a}  ~,~ \gamma_{\b} \,]))_f^{~d} 
~&=~  16 \,  \left[{ {\bm H}}{}^{\mu}{}^{(CS)}\right]{}_{a \, b \, c}{}^d   ~~~~, \cr 
(\gamma^5 \gamma^{\a})_c^{~e}  \,  \left[{ {\bm H}}{}^{\mu}{}^{(CS)}\right]{
}_{a \, b \, e}{}^f  \, (\gamma^5 \gamma_{\a})_f^{~d} 
~&=~ 0~~~~, \cr 
 (\gamma^5)_c^{~e}  \,  \left[{ {\bm H}}{}^{\mu}{}^{(CS)}\right]{}_{a \, b \, e}{}^f  
\, (\gamma^5)_f^{~d} 
~&=~    \left[{ {\bm H}}{}^{\mu}{}^{(CS)}\right]{}_{a \, b \, c}{}^d
~~~~~\,~~,
}   \eqno(A.1)
$$
$$
 \eqalign{
(\gamma^{\a})_c^{~e}  \,  \left[{ {\bm H}}{}^{\mu}{}^{(VS)}\right]{}_{a \, b \, 
e}{}^f  \, (\gamma_{\a})_f^{~d} 
~&=~  +i\, 4 \,C_{ab}
(\gamma^{\m})_c{}^{d} ~-~ i  \, 4 \,(\gamma^{5}
)_{ab} (\gamma^{5}\gamma^{\m})_{c}{}^{d}     \cr
 &~~~~~+i \, 8 \,(\gamma^{5}\gamma^{\m})_{ab}(\gamma^{5})_{c}{
 }^{d}   ~~\,~~~, \cr 
([\, 
\gamma^{\a}  ~,~ \gamma^{\b} \,]))_c^{~e}  \,  \left[{ {\bm H}}{}^{\mu}{
}^{(VS)}\right]{}_{a \, b \, e}{}^f   \, ([\, \gamma_{\a}  ~,~ \gamma_{\b} \,]))_f^{~d} 
~&=~  i  \, 96  \,(\gamma^{5}\gamma^{\m})_{ab}(\gamma^{5})_{c}{
 }^{d}    ~~~~, \cr 
(\gamma^5 \gamma^{\a})_c^{~e}  \,  \left[{ {\bm H}}{}^{\mu}{}^{(VS)}\right]{}_{a \, b \, e}{}^f  
\, (\gamma^5 \gamma_{\a})_f^{~d} 
~&=~  i 4 \,C_{ab}
(\gamma^{\m})_c{}^{d} ~-~ i  \, 4 \,(\gamma^{5}
)_{ab} (\gamma^{5}\gamma^{\m})_{c}{}^{d}     \cr
 &~~~~~-i 8 \,(\gamma^{5}\gamma^{\m})_{ab}(\gamma^{5})_{c}{
 }^{d}      ~~\,~~~, \cr 
 (\gamma^5)_c^{~e}  \,  \left[{ {\bm H}}{}^{\mu}{}^{(VS)}\right]{}_{a \, b \, e}{}^f  
\, (\gamma^5)_f^{~d} 
~&=~  i \, 2 \,C_{ab}
(\gamma^{\m})_c{}^{d} ~+~ i  \, 2 \,(\gamma^{5}
)_{ab} (\gamma^{5}\gamma^{\m})_{c}{}^{d}     \cr
 &~~~~~-i  \, 2 \,(\gamma^{5}\gamma^{\m})_{ab}(\gamma^{5})_{c}{
 }^{d}     
~~~~~,
}  \eqno(A.2)
$$

$$
 \eqalign{
(\gamma^{\a})_c^{~e}  \,  \left[{ {\bm H}}{}^{\mu}{}^{(TS)}\right]{}_{a \, b \, 
e}{}^f  \, (\gamma_{\a})_f^{~d} 
~&=~  -i\, 4 \,C_{ab}
(\gamma^{\m})_c{}^{d} ~-~ i  \, 4 \,(\gamma^{5}
)_{ab} (\gamma^{5}\gamma^{\m})_{c}{}^{d}    \cr
 &~~~~~-i \, 8 \,(\gamma^{5}\gamma^{\m})_{ab}(\gamma^{5})_{c}{
 }^{d}   ~~~~, \cr 
([\, 
\gamma^{\a}  ~,~ \gamma^{\b} \,]))_c^{~e}  \,  \left[{ {\bm H}}{}^{\mu}{}^{(TS)}
\right]{}_{a \, b \, e}{}^f   \, ([\, \gamma_{\a}  ~,~ \gamma_{\b} \,]))_f^{~d} 
~&=~  -\, i \, 96 \,(\gamma^{5}\gamma^{\m})_{ab}(\gamma^{5})_{c}{
 }^{d}    ~~~~~~, \cr 
(\gamma^5 \gamma^{\a})_c^{~e}  \,  \left[{ {\bm H}}{}^{\mu}{}^{(TS)}\right]{}_{
a \, b \, e}{}^f  \, (\gamma^5 \gamma_{\a})_f^{~d} 
~&=~  - i \, 4 \,C_{ab}
(\gamma^{\m})_c{}^{d}  ~-~ i  \, 4 \,(\gamma^{5})_{ab} (\gamma^{5}\gamma^{
\m})_{c}{}^{d}     \cr
 &~~~~~+i  \, 8 \,(\gamma^{5}\gamma^{\m})_{ab}(\gamma^{5})_{c}{
 }^{d}      ~~~~, \cr 
 (\gamma^5)_c^{~e}  \,  \left[{ {\bm H}}{}^{\mu}{}^{(TS)}\right]{}_{a \, b \, e}{}^f  
\, (\gamma^5)_f^{~d} 
~&=~ - i \, 2 \,C_{ab}
(\gamma^{\m})_c{}^{d} ~+~ i  \, 2 \,(\gamma^{5}
)_{ab} (\gamma^{5}\gamma^{\m})_{c}{}^{d}   \cr
 &~~~~~+i  \, 2 \,(\gamma^{5}\gamma^{\m})_{ab}(\gamma^{5})_{c}{
 }^{d}     
~~~~.
}  \eqno(A.3)
$$

\newpage

\end{document}